\documentclass[a4paper]{jpconf}
\usepackage{graphicx}
\begin{document}
\title{Emergence of the stochastic resonance in glow discharge plasma}

\author{Md Nurujjaman$^1$, A N  Sekar Iyengar$^1$ and P Parmananda$^2$}

\address{$^1$Plasma Physics Division, Saha Institute of Nuclear Physics,
1/AF, Bidhannagar, Kolkata -700064, India \\
  $^2$Facultad de Ciencias, UAEM, Avenida Universidad 1001,
Colonia Chamilpa, Cuernavaca, Morelos, M\'{e}xico}

\ead{$^1$jaman\_nonlinear@yahoo.co.in and md.nurujjaman@saha.ac.in}

\begin{abstract}
Stochastic resonance (SR) has been  studied experimentally in a glow discharge plasma. For the SR phenomena, it is observed that a superimposed subthreshold periodic signal can be recovered via stochastic modulations of the discharge voltage.  In the present experiments,  induction of SR is quantified using  the absolute mean difference (AMD). Comparison of the AMD techniques with respect to cross-correlation has also been shown.
\end{abstract}


\section{Introduction}
\label{sec:introduction}
Stochastic resonance (SR) which has been observed in many physical, chemical and
biological systems ~\cite{JPhysA:benzi, arxiv:benzi,revmodphys:Gammaitoni,prl:bruce,pre:parmananda,JStatPhys:moss, prl:longtin,JPhysChem:foster,JPhysChem:Amemiya,prl:kitajo}, is a phenomenon in which the response of the nonlinear system to a weak periodic input signal is amplified or optimized by
the presence of a particular level of noise~\cite{JPhysA:benzi}, i.e.
a previously untraceable subthreshold signal applied to a nonlinear system,
can be detected in the presence of noise. Furthermore, there
exists an optimal level of noise for which the  most efficient detection
takes place. 

\begin{figure*}[ht]
\centering
\includegraphics[width=9cm]{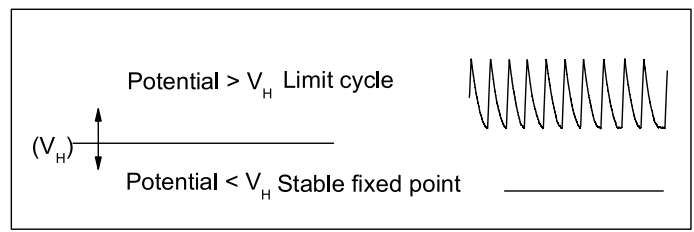}
\caption{It shows that when the control parameter crosses the potential barrier (PB), i.e., $potential >$PB system behaves
shows limit cycle behavior (up) and when ($potential<$ PB) system shows stable behavior (down).}
\label{fig:limit}
\end{figure*}

In 1993, Gang, \emph{et al.}~\cite{prl:gang}, and in 1994, Kurt Wiesenfeld, \emph{et al.}~\cite{prl:wiesenfeld}, had shown  that SR is also possible in different classes of dynamical systems based not on bistability but on excitable dynamics. They proposed a system consisting of a potential barrier (PB) and above this barrier the system shows deterministic dynamics (limit cycle oscillation) and below, stable fixed point as illustrated in Fig~\ref{fig:limit}.
The figure shows that when the control parameter (potential) crosses the PB, i.e., the potential is greater than PB, the system shows a limit cycle behavior and when the potential is less than PB, the system exhibits a stable fixed point behavior.  Now if we set the control parameter (potential) below the PB and apply noise (stochastic perturbation on the potential) to the system and whenever the barrier is crossed, the system returns to its fixed point or ``rest state" deterministically~\cite{prl:gang,prl:wiesenfeld,pre:strogatz}, i.e., whenever the barrier is crossed, the system traverses one oscillation.  Now when a subthreshold periodic signal in the form of pulse and noise are added to the system below the potential barrier the probability of crossing the barrier by noise at the time of occurrence of the periodic pulse is maximum and hence one can get back deterministic dynamics in the form of  periodic oscillations of frequency of the applied periodic pulse for optimum noise level which is the SR of an excitable system. Based upon excitability, SR has been observed in chemical systems~\cite{pre:Santos1}, human brain and many other systems~\cite{revmodphys:Gammaitoni,prl:kitajo}.

In plasma  excitability arises  from the most
fundamental processes, namely the wave-wave and wave-particle interactions.
Different modes may be excited due to nonlinear coupling
of waves and plasma components and the character of the
oscillations is  primarily determined by the plasma parameters and
perturbations~\cite{pr:duncan,pr:sturrock,pop:Shokri}.  In our experiments,  we get excitable dynamics, for certain discharge parameters, in the region greater than the Paschen minimum~\cite{chaos:Nurujjaman} and has been discussed in Section~\ref{section:autonomous}.
\section{Experimental setup}
\label{section:Experimental setup }
\begin{figure}[ht]
\centering
\includegraphics[width=8.5cm]{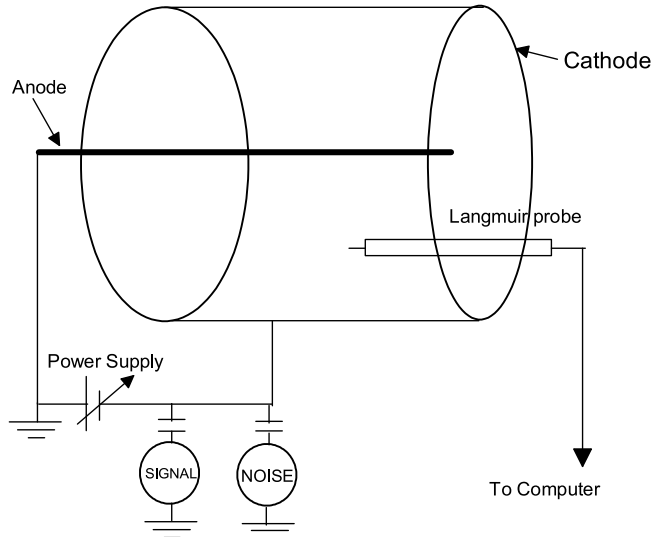}
\caption{Schematic diagram of the cylindrical
electrode system of the glow discharge plasma.
The probe was placed at a distance $l \approx12.5$ mm
from the anode. Signal and noise sources were
coupled to the discharge voltage through a capacitor.}
\label{fig1:setup}
\end{figure}

The experiments were performed in a hollow cathode
dc glow discharge plasma. The schematic diagram of
the experimental setup  is presented in Fig~\ref{fig1:setup}.
A hollow stainless steel tube of length $\approx7~cm$ and of
diameter ($\phi$) $\approx$ 45 mm was used as the cathode and
a central rod of length $\approx7~cm$ and $\phi\approx$ 1.6 mm was employed
as the anode. The whole assembly was mounted inside a
vacuum chamber and was pumped down to a pressure
of  about  0.001 mbar using a rotary
pump. The chamber was subsequently filled with the
argon gas up to a pre-determined  value of neutral
pressure by a needle valve. Finally  a discharge was struck
by a dc discharge voltage (DV), which could be varied in
the range of 0$-$1000 V.

The noise and subthreshold periodic square pulse generators were
coupled with DV through a capacitor [Fig.~\ref{fig1:setup}]. In
all the experiments DV was used as  the bifurcation parameter
while the remaining system parameters like pressure etc.,
were maintained constant. The system observable
was the electrostatic floating potential, which
was measured using a Langmuir probe of
diameter $\phi$ = 0.5 mm  and  length
2 mm. The tip of this Langmuir probe was placed in
the center of the electrode system as indicated in
Fig.~\ref{fig1:setup}.
The  plasma density and the electron
temperature were determined to be of the
order of 10$^7$cm$^{-3}$ and 3$-$4 eV respectively.
Furthermore, the  electron plasma frequency ($f_{pe}$)
was estimated to be around  28 MHz, whereas the ion plasma
frequency ($f_{pi}$) was estimated to be around  105 kHz.
\section{Autonomous dynamics}
\label{section:autonomous}
\begin{figure}[ht]
\centering
\includegraphics[angle=-90,width=13.5cm]{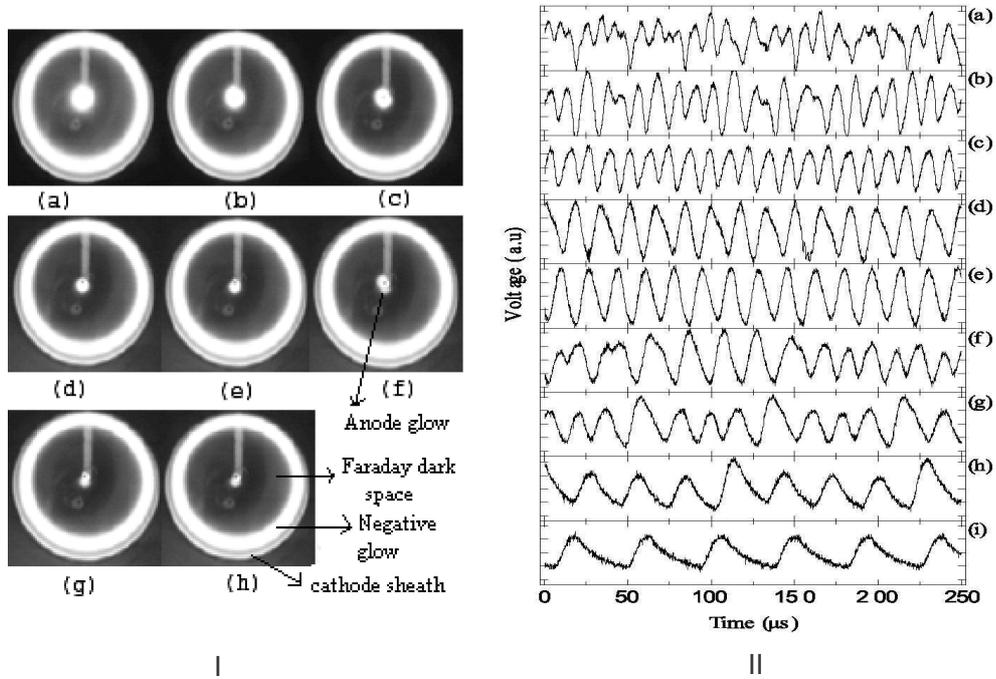}
\caption{I. Evolution of glow size of the anode glow with increasing DV (a) - (h). II. Sequential change in the raw signal (normalized) at 0.95 mbar for different voltages:(a) 283 V; (b) 284 V; (c) 286 V; (d) 288 V; (e) 289 V; (f) 290 V; (g) 291 V; (h) 292 V; (i) 293 V. All y-axes range form -1 to 1.}
\label{fig:auto}
\end{figure}

At high pressures an anode glow and fluctuations in the floating potential were simultaneously observed as shown in Figs.~\ref{fig:auto}(I) and \ref{fig:auto}(II) respectively. Figs ~\ref{fig:auto}I(a) shows that the glow with largest size, appears when the discharge is struck at a typical pressure of 0.95 mbar and its size decreases with increase in the DV until it finally disappears [Figs~\ref{fig:auto}I(a)$-$\ref{fig:auto}II(h)]. 
From the CCD image analysis the annular radius of the glow around the anode, was estimated to be $\approx 1.3$ mm at the beginning of the discharge [Fig~\ref{fig:auto}I(a)] and reduced to $\approx 0.32$ mm [Fig~\ref{fig:auto}I(f)]. Based on the observation of Valentin Pohoa\c{t}\v{a}, \emph{et al.}~\cite{pre:Valentin}, we feel that the relaxation oscillations are double layers or such coherent potential structures, which are constantly forming and annihilating. An interesting feature associated with the anode glow was the different types of oscillations in the floating potential at different pressures. The fluctuations in the floating potential fluctuations were irregular at the initial stage of discharge [Fig~\ref{fig:auto}II(a)] and became regular with  increase in the DVs Figs~\ref{fig:auto}II(h)$-$\ref{fig:auto}II(i), and finally reaches a stable fixed point through  homoclinic bifurcation~\cite{pre:nurujjaman}. The DV at which these oscillations cease may be termed as the bifurcation point ($V_H$) which acts as a PB in this experiment. The floating potential fluctuation exhibits relaxation oscillations on the one side of the $V_H$ and stable fixed point on the other side,  which is termed as being in an excitable state and is useful to study SR.

\section{Stochastic resonance}
\label{section:SR}
For our experiments on SR, the reference
voltage $V_0$ was chosen such that $V_0>V_H$ and therefore
the autonomous dynamics, by virtue of
an underlying homoclinic bifurcation, exhibit
steady state behavior. The DV
was thereafter perturbed  $V=V_0+S(t)+D\xi$,
where  $S(t)$ is the subthreshold periodic pulse train
chosen for which $V=V_0+S(t)>V_H$, (subthreshold
signal does not  cause the  system to cross over
to the oscillatory regime) and $D\xi$
is  the added Gaussian white noise $\xi$ with
amplitude  $D$.
 \begin{figure*}[ht]
\centering
\includegraphics{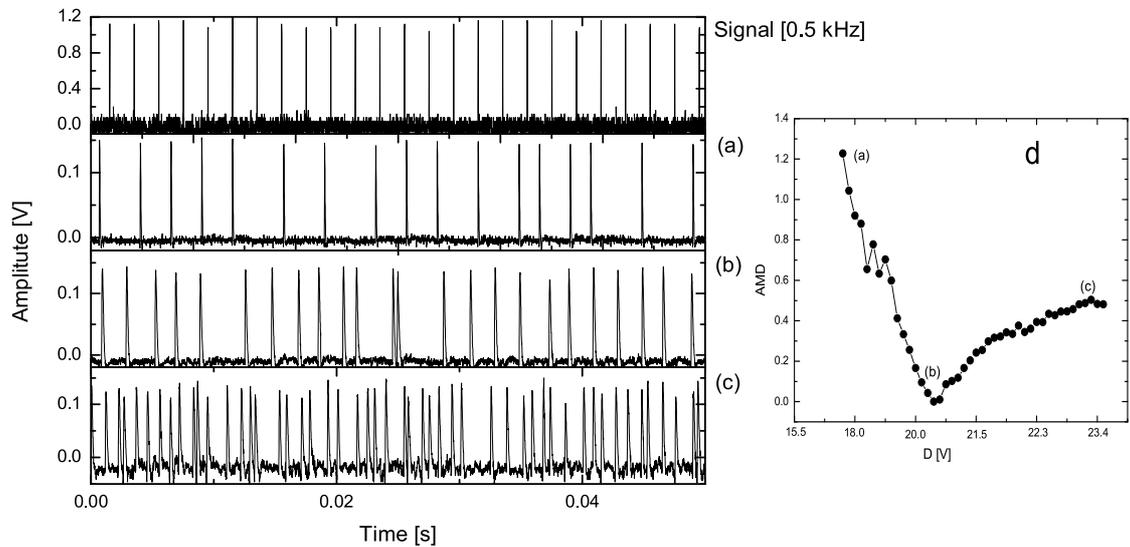}
\caption{Noise induced system response in the floating potential
fluctuations for low, medium , and high amplitude noise in
conjunction with a subthreshold periodic square pulse. The right
panel [Fig~\ref{fig:periodic}] shows the AMD as a function of
noise amplitude for the experiment performed at $V_0=307$ V and
pressure= 0.39 mbar. Left panel shows the subthreshold periodic
pulse train and the  three time series of floating
potential fluctuations at low level noise (a); at optimum
noise value (b) and at high amplitude noise (c).}
\label{fig:periodic}
\end{figure*}
 Subthreshold periodic square pulse of
width $20~\mu s$ and duration 2 ms was constructed
using Fluke PM5138A function generator. Meanwhile,  the
gaussian noise  produced using the HP 33120A noise
generator was subsequently amplified using a noise
amplifier.

Fig.~\ref{fig:periodic}$(a)-(c)$ show time
series of the system response  in the
presence of an identical subthreshold
signal for three different
amplitudes of imposed noise. The subthreshold
periodic pulse train is also plotted, in the
top most graph of the left panel, for comparison
purposes. Fig.~\ref{fig:periodic}(a) shows
that there is little correspondence between the subthreshold
signal and the system response for a low noise amplitude.
However, there is excellent correspondence at an intermediate noise
amplitude [Fig.~\ref{fig:periodic}(b)]. Finally,  at higher
amplitudes of noise the subthreshold signal is lost amidst
stochastic fluctuations of the system
response [Fig.~\ref{fig:periodic}(c)]. Absolute
mean difference (AMD),  used to quantify the
information transfer between the subthreshold
signal and the system response, is defined
as $AMD=abs(mean(\frac{t_p}{\delta}-1))$.
$t_p$ and $\delta$ are the inter-peak
interval of the response signal
and mean peak interval of the subthreshold periodic
signal respectively.  Fig~\ref{fig:periodic}(d) shows
that the experimentally computed AMD versus noise amplitude D curve
has a unimodal structure typical for the SR phenomena.
The minima in this curve  corresponds to
the optimal noise level for which maximum information transfer
between the input and the output takes place.

\section{\label{sec:3}Discussion}
The effect of noise has been studied experimentally near
a homoclinic bifurcation in glow discharge plasma
system. Our study demonstrates the  emergence of SR for
periodic subthreshold square pulse signals  via purely stochastic fluctuations.

In SR experiments, the efficiency of information transfer
was quantified using AMD instead of the  power
norm which has been utilized elsewhere ~\cite{pre:parmananda}.
The advantage of using  this method in comparison to  the power
norm ($C_0(0)$)~\cite{pre:parmananda} lies in
the fact that AMD remains independent of the lag between the
measured floating potential and the applied periodic
square pulse. This is of relevance to our experimental
system, where  invariably there exists  a lag, at times
varying in time due to the parameter drifts.
\begin{figure}[htp]
\centering
\includegraphics[width=8.5cm, height=8cm]{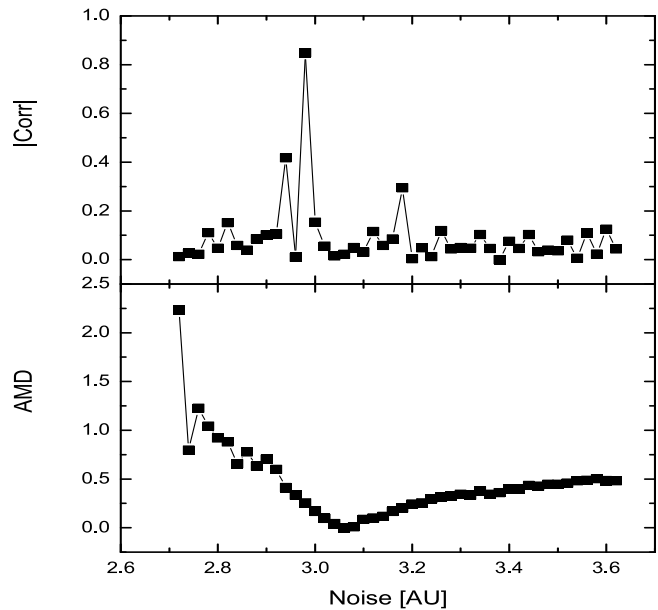}
\caption{Upper and lower panels show $C_0(0)$ vs applied noise amplitude and corresponding AMD respectively. $C_0(0)$ plot does not show any maximum at optimum noise but  AMD shows clear minimum at optimum noise level.}
\label{fig:crAMD}
\end{figure}
Comparison between the estimated $C_0(0)$ and AMD
have been shown in Fig~\ref{fig:crAMD}. It is obvious from the upper panel of Fig~\ref{fig:crAMD} that $C_0(0)$ does not show any peak
at optimum noise level. Whereas, AMD shows nice agreement with regular spiking of the signal [Fig~\ref{fig:crAMD}(lower panel).
Derivation of the AMD statistics has been shown in ~\ref{appendix:amd}.

\ack

The authors acknowledge
A. Bal, S.S. Sil, A. Ram and D. Das and the Micro Electronic Division
of SINP for their technical help during the experiments.

\appendix

\section{Absolute mean difference (AMD)}
\label{appendix:amd}
Absolute mean difference (AMD) is the statistical tool, proposed to quantify the SR in a plasma subjected to noise and a periodic signal. AMD is defined as $AMD=abs(mean(t_p/\delta-1))$, and gives  AMD gives the degree of mimicking the output to the input subthreshold signal.
Usually, regularity in the stochastic resonance is quantified
by calculating cross-correlation ($Co = | < [(x1- < x1 >
)(x2- < x2 >)] > |$) between the output and input signal. But in case of plasma it is not suitable,
because there is always a lag between periodic signal that is applied to the plasma and
the output. This lag also varies with time because the plasma
conditions keep fluctuating over time. Therefore,
cross-correlation is not the right quantity to be estimated. So we have proposed a statistical tool AMD which will be independent of lag and is estimated as follows:
\begin{enumerate}
\item First calculate the mean inter-peak distance ($\delta$) of the
periodic signal.
\item Calculate the inter-peak distances ($t_p$) of the output
signal.
\item Calculate the ($(t_p -\delta)/\delta$)
\item Take the absolute, i.e.,    $| < ( t_p/\delta - 1) > |$.
\end{enumerate}
Therefore, $AMD=abs(mean(\frac{t_P}{\delta}-1))$.

\end{document}